\documentclass[12pt]{article}

\usepackage[top=1.25in, bottom=1.25in, left=1.25in, right=1.25in]{geometry}
\usepackage{setspace}
\usepackage[hang,flushmargin]{footmisc} 

\usepackage{amsmath}      
\usepackage{amssymb}      
\usepackage{mathtools}    
\usepackage{mathrsfs}     
\usepackage{bm}           
\usepackage{dsfont}       
\usepackage{bbm}          
\usepackage{amsthm}

\usepackage[dvipsnames, table]{xcolor}
\usepackage{graphicx}
\usepackage{tikz}
\usepackage{pgfplots}
\pgfplotsset{compat=1.18} 

\usepackage{array}
\usepackage{float}
\usepackage{multirow}
\usepackage{tabularx}
\usepackage{caption}
\usepackage{subcaption}
\usepackage{longtable}
\usepackage{booktabs}

\usepackage{thmtools}     
\usepackage{enumitem}     

\usepackage{natbib}
\usepackage{bibentry}

\usepackage{appendix}
\usepackage{datetime}
\usepackage{textcase,relsize}
\usepackage{xr}           

\usepackage[colorlinks,citecolor=blue,urlcolor=blue,hypertexnames=false]{hyperref}

\allowdisplaybreaks 

\newtheoremstyle{named-normal}{}{}{\normalfont}{}{\bfseries}{.}{.5em}{#1 (\thmnote{#3})}
\newtheoremstyle{named-italic}{}{}{\itshape}{}{\bfseries}{.}{.5em}{#1 (\thmnote{#3})}
\declaretheorem[style=named-normal,numbered=yes,name=Assumption]{named-asmp}
\declaretheorem[style=named-normal,numbered=yes,name=Condition]{named-cond}
\declaretheorem[style=named-normal,numbered=yes,name=Example]{named-exmp}

\newdateformat{monthyeardate}{\monthname[\THEMONTH], \THEYEAR}

\def \bbB {\mathbb{B}}

\def \bbE {\mathbb{E}}

    \def \calL {\mathcal{L}}

\def \bbP {\mathbb{P}}    \def \calP {\mathcal{P}}

    \def \calS {\mathcal{S}}

    \def \calW {\mathcal{W}}

\newcommand{\msone}{\mspace{1mu}}

\def \Var {\text{Var}}

\usepackage{setspace}
\usepackage{lipsum}
\usepackage{accents}

\usepackage{aurical}
\usepackage[T1]{fontenc}
\newcommand{\psk}{\text{\Fontskrivan{p}}}
\newcommand{\qsk}{\text{\Fontskrivan{q}}}

\newcommand{\ATE}{\mathrel{\scriptscriptstyle \texttt{ATE}}}

\newcommand{\IPW}{\mathrel{\scriptscriptstyle \texttt{IPW}}}

\newcommand{\Reg}{\mathrel{\scriptscriptstyle \texttt{Reg}}}
\newcommand{\LR}{\mathrel{\scriptstyle \texttt{LR}}}

\makeatletter
\def\thm@space@setup{%
  \thm@preskip=1.5\topsep
  \thm@postskip=\thm@preskip 
}
\makeatother

\numberwithin{equation}{section}
\theoremstyle{plain}
\newtheorem{theorem}{Theorem}[section]

\newtheorem{assumption}{Assumption}[section]

\newtheorem{definition}[theorem]{Definition}

\newtheorem{lemma}[theorem]{Lemma}

\theoremstyle{definition}
\newtheorem{example}{Example}

\def \Frechet {Fr\'{e}chet}
\def \Gateaux {G\^{a}teaux}

\def \vec {\text{vec}}

\newcommand{\Pon}[1]{P_{(:|{#1})}}

\newcommand{\Peg}[1]{P_{({#1})}}

\definecolor{ao}{rgb}{0.0, 0.5, 0.0}

\title{Influence Functions: A Unified Differentiation-Based Framework}
\author{Xiye Yang and Ruonan Xu \\[0.3em] Department of Economics\\ Rutgers University}
\date{\monthyeardate\today} 

\begin{document}
\bibliographystyle{ecta}

\maketitle
\setlength{\baselineskip}{1.0\baselineskip}
\hyphenpenalty=5000 \tolerance=1000

\begin{abstract}
\singlespace

This paper introduces a direct differentiation-based framework that unifies the derivation of influence functions across parametric, nonparametric, and semiparametric models. We show that the Riesz representer of the functional derivative is obtained by orthogonally projecting the identification function onto the subspace of mean-zero functions, so the influence function is a linear transformation of the moment function. This approach extends seamlessly to infinite-dimensional parameters and reveals a common algebraic form across both finite- and infinite-dimensional cases. It demonstrates that the slower nonparametric convergence rate arises from approximating the infeasible Dirac delta function. The same framework further applies to semiparametric multi-step plug-in estimation, where it automatically yields locally robust moment functions. Crucially, by employing distributional derivatives, the method accommodates non-smooth moment conditions without imposing classical differentiability.

\medskip


\end{abstract}

\section{Introduction}

The influence function is a foundational object in modern statistics and econometrics, characterizing the first-order effect of an infinitesimal perturbation to the data-generating distribution on a target parameter or estimator. It plays a central role in establishing asymptotic linearity, semiparametric efficiency, and local robustness: seminal work by \cite{BKRW:1993} uses it to characterize semiparametric efficiency bounds, while recent literature \citep[e.g.,][]{CCDDHN:2018double,LocalRobust:2022} highlights its role in constructing locally robust moment functions.

Despite its theoretical centrality, deriving influence functions remains technically demanding. Stochastic expansions of the estimation error are often algebraically cumbersome. While tangent-space method \citep{tsiatis2006semiparametric} is mathematically elegant, it can be abstract and opaque for complex models. As \cite{Hahn:1998} notes, the general frameworks of \cite{Newey:1990, Newey:1994} and \cite{BKRW:1993} may require one to posit a candidate influence function and verify its properties ex post.  \cite{Hahn&Ridder:2013} extend \cite{Newey:1994} to three-step settings but restrict the analysis to differentiable functions. The method by \cite{Ichimura&Newey:2022} ultimately require solving a least square projection problem, where the nuisance moment function is scalar-valued. A unified, explicit, and calculus-driven recipe remains absent.

This work establishes a cohesive analytical framework, grounded in functional differentiation and chain rule, for the unified derivation of influence functions across parametric, nonparametric, and semiparametric cases. We define the target parameter as a statistical functional of the probability distribution $P$ and compute its functional derivative under local perturbations of the form $P^\epsilon \!=\! (1-\epsilon)P + \epsilon Q$. For regular parameters, we show that the Riesz representer of this derivative is obtained by orthogonally projecting the identification function onto the subspace of mean-zero square-integrable functions. Crucially, this projection reduces to a simple centering operation: subtracting the unconditional or conditional expectation from the identification function (see Section~\ref{subsec:DirectInd}).

The framework extends seamlessly to infinite-dimensional parameters. While the convergence rates differ fundamentally between finite- and infinite-dimensional settings, we demonstrate that the influence functions share a common algebraic structure: in both cases, they arise as linear transformations of the moment function (see Section~\ref{sec:Nonparametric}). For finite-dimensional parameters, the transformation operator is a constant matrix. For infinite-dimensional parameters, it becomes an approximation to the infeasible Dirac delta function and has diverging variance, which slows down the rate of convergence. To make this concrete, we provide multiple application strategies and distill the derivation into chain-rule type results (Lemmas~\ref{lem:Key-Uncond} and ~\ref{lem:Key-Cond}). 

In semiparametric multi-step plug-in estimation, our approach immediately yields the locally robust moment function. Deriving the associated adjustment term reduces to computing conditional expectations of partial derivatives. Crucially, this procedure does not require classical differentiability. By employing distributional derivatives, the method accommodates non-smooth moment conditions, with the conditional expectation serving as a smoothing operator.

This paper is organized as follows. Section~\ref{sec:Definition} introduces the formal setup and key definitions. Section~\ref{sec:Parametric} illustrates how our differentiation-based method works in some prototype parametric examples and gives a simple chain rule in the case of unconditional distributions (Lemma~\ref{lem:Key-Uncond}). Section~\ref{sec:Nonparametric} discusses nonparametric influence function and extends the chain rule to the conditional case. The most general results are summarized in Section~\ref{sec:Semiparametric}.
Section~\ref{sec:conclusion} concludes. All technical proofs and supplementary results are collected in the online supplement.

\section{Notation and Definitions} \label{sec:Definition}

\subsection{Parameter: identification and estimation}

Let $Z$ be a random vector with distribution $P$. Following \cite{BKRW:1993}, all parameters are viewed as functionals of $P$, e.g., $\nu\!=\!\nu(P)$ for a generic parameter $\nu$. This makes it more straightforward to define its influence function $\dot{\nu}$ as a functional derivative in the next subsection. Expectations are written as $P[f(Z)] \!\coloneqq\! \int f(z) dP(z)$, where $z$ is a real vector. Whenever possible, we will omit $Z$ or $z$ for simplicity.

We consider two types of identification for the parameter of interest $\beta$:
\begin{align} 
	\text{Direct: }& \beta(P) = P[h_\beta(\gamma(P))] = P\big[h_\beta\big(\gamma_1(P), \gamma_2(\Pon{1}), \ldots, \gamma_l(\Pon{l-1}) \big)\big], \label{eq:betaDirect} \\
	\text{Indirect: }& P[m_\beta(\beta, \gamma(P))] = P\big[m_\beta\big(\beta, \gamma_1(P), \gamma_2(\Pon{1}), \ldots, \gamma_l(\Pon{l-1}) \big)\big] = 0, \label{eq:betaInDirect} 
\end{align}
where $h_\beta$ and $m_\beta$ are known functions and $\gamma = (\gamma_1^\intercal, \ldots, \gamma_l^\intercal)^\intercal$ is the vector containing all the nuisance parameters. The direct case is a special instance of the indirect one, upon noticing that one can set $m_\beta \!=\! h_\beta - \beta$. When there are no nuisance parameters, an estimator of $\beta$ can be obtained by simply replacing $P$ with the empirical measure $\bbP_n$, i.e., $\hat{\beta}\!=\!\beta(\bbP_n)$. The asymptotic properties of $\hat{\beta}$ then readily follow from the well-established empirical process theory \citep{EmpiricalProcess}.

Each $\gamma_j$ is assumed to be identified from $\Pon{j-1}[m_{\gamma_j}]\!=\!0$, which may or may not reduce to a direct identification. The conditional distribution $\Pon{j-1}$ is associated with the decomposition $Z \!=\! (Z^{(1)\intercal}, \ldots, Z^{(l)\intercal})^\intercal$. For $i \leq j$, let $Z^{(j:i)} \!=\! (Z^{(i)\intercal}, \ldots, Z^{(j)\intercal})^\intercal$. For $j \!=\! 2, \ldots, l$, let $\Pon{j-1} \!\coloneqq\! P_{(l:j|j-1:1)}$ denote the conditional distribution of $Z^{(l:j)}$ given $Z^{(j-1:1)}$. When $j\!=\!1$, $\Pon{0} \!=\! P$ reduces to the unconditional distribution. The decomposition is based on the levels of ``exogeneity/endogeneity'' of each $Z^{(j)}$. For example, we have $P_{(2|1)} \!=\! P_{(Y|X)}$ in linear regression model. In instrumental variables (IV) estimation, $Z^{(1)} \!=\! W$ consists of all the instruments and exogenous variables and $Z^{(2)} \!=\! (Y,X^\intercal)^\intercal$, where $X$ only contains endogenous variables. In treatment effect analysis, $P_{(:|2)} \!=\! P_{(Y|T, X)}$ and $P_{(:|1)} \!=\! P_{(Y,T|X)}$, where $T$ is the treatment indicator. 

In econometric literature, conditional and unconditional moment restrictions, as well as finite- and infinite-dimensional parameters, are often treated separately. Here, we try to find some symbolic similarities by adopting a generalized function perspective. More specifically, define the Dirac delta function $\delta_z$ as the distributional derivative \citep{Jones:1982, Hormander:1983} of the Heaviside function $H_z(z') \!=\! 1_{\{z' \geq z\}}$, which satisfies $H_z[f] \!=\! f(z)$ in the Lebesgue–Stieltjes integral sense. With some abuse of notation, we symbolically write $H_z[f] \!=\! \int f(z') \delta_z(z') \mu(dz')$. Let $Z^{(1)}=X$ and consider the identification of the value of an unknown function $\gamma$ at $x$ (denoted by $\gamma_x$):
\begin{align} \label{eq:Momdelta}
	P[m_\gamma(Z, \gamma) | X = x] =  0  \,\, \Longleftrightarrow \,\, \frac{P[m_\gamma(Z, \gamma) \delta_{x}]}{P[\delta_{x}]} = 0 \,\, \Longleftrightarrow \,\, P[m_\gamma(Z, \gamma) \delta_x^\dag] = 0,
\end{align}
where $\delta_{x}^\dag \!\coloneqq\! P[\delta_{x}]^{-1} \delta_{x}$. It will become more transparent in Section~\ref{sec:Nonparametric} that different nonparametric methods amount to different approximations to this generalized function. Note that the introduction of the generalized function $\delta_x^\dag$ symbolically transforms the conditional moment condition into an unconditional one.

%

When \eqref{eq:Momdelta} holds for all $x$ in the support of $X$, we can obtain a conditional moment condition $P[m_\gamma(Z, \gamma) \!\mid\! X] \!=\! 0$, without the appearance of generalized functions. Intuitively, this identifies the infinite-dimensional parameter $\gamma(\cdot)$, which is a function of $x$. It is also a statistical functional of the conditional distribution $P_{(Y|X)}$. We show in Section~\ref{sec:Nonparametric} that computing the influence function of $\gamma(\cdot)$, denoted by $\dot{\gamma}$, is significantly more straightforward than computing $\dot{\gamma}_x$, where $\gamma_x \!\coloneqq\! \gamma(x)$. More importantly, the estimator of $\beta$ typically requires the entire function $\gamma(\cdot)$ as input. Consequently, it is $\dot{\gamma}$ rather than $\dot{\gamma}_x$ that appears in the influence function of $\beta$ (see Section \ref{sec:Semiparametric}). 

The above symbolic similarity between finite- and infinite-dimensional parameters turns out to be more substantive. We will show step by step how our differentiation-based methods work across parameteric, nonparametric, and semiparametric cases, with the help of chain rule for functional derivatives.

Throughout the paper, we maintain the strong identification assumption and defer the analysis of weak identification to future work.

\begin{assumption}[Strong Identification] \label{asmp:strong}
Assume there exist positive semi-definite weighting matrices $\calW_{\beta\beta}$ and $\calW_{j}(z^{(j-1:1)})$ such that
\begin{align*}
	P[\partial_\beta m_\beta]^\intercal \msone \calW_{\beta\beta} \msone P[\partial_\beta m_\beta] \quad \text{and} \quad \Pon{j-1}[\partial_{\gamma_j} m_{\gamma,j}]^\intercal \msone \calW_{j}(Z^{(j-1:1)}) \msone \Pon{j-1}[\partial_{\gamma_j} m_{\gamma,j}]
\end{align*}
are invertible almost surely for all $j=1,\ldots,l$. 
\end{assumption}

We require only positive semi-definiteness of the weighting matrices to accommodate potential over-identification of $\beta$ and/or the nuisance parameters $\gamma_j$.

\subsection{Influence function: functional derivative}

Motivated by the structural parallels between finite- and infinite-dimensional parameters, we analyze a generic statistical functional $\nu\!=\!\nu(P)$. Let $\mathcal{L}_2(P)$ denote the space of square-integrable functions with respect to $P$, and let $\mathcal{L}_2^0(P) \!\subseteq\! \mathcal{L}_2(P)$ be the subspace of mean-zero functions. We equip $\mathcal{L}_2(P)$ with the inner product $\langle \cdot, \cdot \rangle$. Following standard semiparametric theory, we define functional derivatives via pathwise differentiability \citep{Hampel:1974, Huber:1984, BKRW:1993}.

\begin{definition}[Directional Derivative]\label{def:derivative}
The directional derivative $\dot{\nu}(P; Q-P)$ of a generic parameter $\nu$ at $P$ along the direction $Q-P$ is defined as
\begin{align} \label{eq:Riesz}
	\dot{\nu}(P; Q-P) \coloneq \frac{d} {d \epsilon} \nu\big( P^\epsilon \big) \big|_{\epsilon=0} = \frac{d} {d \epsilon} \nu\big( P+ \epsilon (Q-P) \big) \big|_{\epsilon=0}.
\end{align}
Let $\calS$ be the collection of directions $Q-P$ such that the above derivative exists. Then different types of functional derivatives (\Gateaux, Hadamard, or \Frechet) correspond to different structures of the set $\calS$ (refer to Appendix and Section A.5 of \cite{BKRW:1993} for more details).\footnote{We do not specify the type of functional differentiability mainly because the asymptotic negligibility of the remainder terms is guaranteed by the stochastic equicontinuity condition \citep{Andrews:1994, Newey:1994, Newey&McFadden:1994}.}
\end{definition}

The influence function is defined similarly across methods, but the challenge lies in deriving its explicit form. For example, \cite{Newey:1994} uses a score-based expression, which is extended to three-step settings by \cite{Hahn&Ridder:2013}; \cite{tsiatis2006semiparametric} relies on tangent space methods; and \cite{Ichimura&Newey:2022} propose a least squares-based approach. 

Our method begins with the integral representation of the directional derivative $\dot{\nu}(P; Q - P)$. To ensure tractability, it is typically assumed that this derivative is linear and continuous in $Q - P$. Under \Gateaux~differentiability, \cite{Huber:1984} (Chapter 2.5) shows that there exists a measurable function, denoted as $\dot{\nu}(z; P)$, such that:
\begin{align} \label{eq:IntRep}
	\dot{\nu}(P; Q-P) = \int \dot{\nu}(z; P) \, d(Q-P)(z) = \int \dot{\nu}(z; P) \, dQ(z),
\end{align}
or, equivalently, in the form of expectations:
\begin{align*}
	\dot{\nu}(P; Q-P) = (Q-P)[\dot{\nu}(Z;P)] = Q[\dot{\nu}(Z;P)].
\end{align*}
where the second equality holds if we restrict $\dot{\nu}(\cdot; P) \in \calL_2^0(P)$ (i.e., $P[\dot{\nu}(Z; P)] = 0$). Without this mean-zero constraint, infinitely many functions could satisfy the first equality (e.g., $\dot{\nu}(z; P) + c$ for any constant $c$).

By the Riesz representation theorem (e.g., \cite{Rudin:1987}), this mean-zero function $\dot{\nu}(\cdot; P)$ is unique and is precisely the influence function (or influence curve) introduced by \cite{Hampel:1968, Hampel:1974}. In essence, the Riesz representation here acts as a simple projection from $\calL_2(P)$ onto $\calL_2^0(P)$ by subtracting the mean. While \cite{BKRW:1993} (p.~19) suggest evaluating this function by taking $Q$ as a point mass at $z$, we will take an alternative approach that applies to any admissible measure $Q$.

\begin{definition}[Influence Function (Functional Derivative)] \label{def:IF}
The influence function (or functional derivative) of $\nu$ at $P$, denoted as $\dot{\nu}(\cdot; P)$ or $\frac{\delta \nu}{\delta P}(\cdot)$, is defined as the unique element of $\mathcal{L}_2^0(P)$ satisfying
\begin{align} \label{eq:IF-def}
	\dot{\nu}(P; Q-P) = (Q-P)[\dot{\nu}(Z; P)] \quad \text{for all admissible } Q-P \in \calS.
\end{align}
\end{definition}

As a comparison, the directional derivative $\dot{\nu}(P; Q-P)$ is a vector representing the rate of change along a specific direction, while the functional derivative $\dot{\nu}(z; P)$ is a function that encodes the sensitivity of $\nu(P)$ to an infinitesimal point-mass contamination at point $z$.

\subsection{Regularity}

The regularity of parameters is closely related to the directional derivative $\dot{\nu}(P; Q-P)$. The following definition is originally applicable to finite-dimensional parameters. However, a slight modification (see Section \ref{subsec:NonPar}) extends this to infinite-dimensional parameters.

\begin{definition}[Regular Parameter] \label{def:regularity}
A parameter $\nu(P)$, mapping $\calP \rightarrow \bbB$ (where $\bbB$ is a Banach space), is regular at $P \in \calP$ if:

(i) Existence of Influence Function: The influence function $\dot{\nu}(\cdot; P) \in \calL_2^0(P)$ exists (Definition \ref{def:IF}), so that the directional derivative satisfies
\begin{align*}
	\dot{\nu}(P; Q-P) = (Q-P)[\dot{\nu}(Z; P)] \quad \text{for all } Q-P \in \calS.
\end{align*}

(ii) Square Integrability: $\dot{\nu}(\cdot; P) \in \calL_2(P)$, i.e., $P[\|\dot{\nu}(Z; P)\|^2] < \infty$.

(iii) Non-degeneracy: The variance-covariance matrix $\Sigma_\nu \coloneqq P[\dot{\nu}(Z; P) \dot{\nu}(Z; P)^\intercal]$ is invertible.

If $\nu$ is regular at all $P \in \calP$ and the map $P \mapsto \dot{\nu}(\cdot; P)$ is continuous in a suitable sense, $\nu$ is termed a regular parameter.
\end{definition}

Intuitively, condition (i) ensures that the functional $\nu$ varies smoothly with respect to $P$, and the existence of an influence function guarantees the possibility of constructing asymptotically linear estimators. Condition (ii) ensures finite asymptotic variance for estimators achieving the parametric $\sqrt{n}$-rate; for nonparametric estimators with slower convergence rates, this integrability requirement needs to be modified to reflect the rate change (see Section \ref{sec:Nonparametric}). Condition (iii) guarantees that the asymptotic covariance matrix is non-degenerate.

\section{Prototype Parametric Examples} \label{sec:Parametric}


\subsection{Direct Identification} \label{subsec:DirectInd}

A standard example of direct identification is $\nu(P) \!=\! P[h(Z)]$, where $h \in \mathcal{L}_2(P)$ is a known function. Because $\nu$ is linear in $P$, its influence function follows directly from projecting $h$ onto the mean-zero subspace $\mathcal{L}_2^0(P)$:
\begin{align} \label{eq:DirectIF}
	\dot{\nu}(P; Q-P) = (Q-P)[h] \;\; \Longrightarrow \;\; \dot{\nu}(Z, P) = h(Z) - P[h(Z)] = h(Z) - \nu(P). 
\end{align}

The core of our approach is to express the directional derivative $\dot{\nu}(P; Q - P)$ in the form $(Q - P)[h]$ for some integrable function $h$. The influence function is then recovered by centering $h$, i.e., subtracting $P[h]$.

This representation parallels the relationship between a directional derivative $\nabla_{\mathbf{v}} f(x) \!=\! \nabla f(x) \cdot \mathbf{v}$ and the gradient $\nabla f(x)$. When $Q$ is absolutely continuous with respect to $P$ with density $\qsk \!=\! dQ/dP$, we have $\dot{\nu}(P; Q - P) \!=\! \langle \dot{\nu}(Z,P), \qsk \rangle$, where $\dot{\nu}(\cdot,P)$ acts as the functional gradient. As in finite-dimensional calculus, identifying the inner product of $\dot{\nu}(\cdot,P)$ with a sufficiently rich class of perturbation densities uniquely determines the influence function, even in infinite-dimensional settings. We will extend this identification argument below in the context of sequential plug-in estimation under both unconditional and conditional constraints.

\subsection{Indirect identification}\label{subsec:IndirectInd}

Consider a parameter $\nu$ identified indirectly via the unconditional moment restriction
\begin{align*}
	P[m(Z, \nu)] = 0,
\end{align*}
where $d_m \geq d_\nu$ so that $\nu$ could be potentially over-identified. Although the implicit function $\nu(P)$ typically lacks a closed-form expression, its influence function $\dot{\nu}(Z; P)$ admits an explicit analytic representation.

Let $A$ be a $d_\nu \times d_m$ matrix of full row rank, and assume $A \, P[\partial_\nu m(Z)]$ is invertible, which constitutes a strong identification condition. Now consider the transformed moment condition $P[A \, m(\nu)] = 0$ and the perturbation $P^\epsilon = P + \epsilon(Q-P)$ along the direction $Q$. Let $\nu^\epsilon \coloneqq \nu(P^\epsilon)$ solve the perturbed moment condition,
\begin{align*}
	P^\epsilon[A \, m(Z,\nu^\epsilon)] = 0
\end{align*}
for all $\epsilon$ in a neighborhood of $0$.

Since $\nu$ is a regular parameter, the map $\epsilon \mapsto P^\epsilon[A\,m(\nu^\epsilon)]$ is differentiable at $\epsilon = 0$. Differentiating the identity with respect to $\epsilon$ and evaluating at $\epsilon = 0$ gives
\begin{align} \label{eq:diff-identity}
	0 = \frac{d}{d\epsilon} P^\epsilon[A\,m(Z, \nu^\epsilon)] \Bigm|_{\epsilon=0}
	= \underbrace{(Q-P)[A\,m(Z,\nu)]}_{\text{measure perturbation}} + \underbrace{\frac{d}{d\epsilon} P[A\,m(Z,\nu^\epsilon)] \Bigm|_{\epsilon=0}}_{\text{parameter perturbation}}.
\end{align}
The measure perturbation term uses only continuity of the integrated map $\epsilon \mapsto (Q-P)[A\,m(\nu^\epsilon)]$ at $\epsilon \!=\! 0$. The parameter perturbation term demands more detailed discussion.

When $\nu \mapsto m(z,\nu)$ is differentiable, one can simply interchange the order of differentiation and expectation and get
\begin{align*}
	\frac{d}{d\epsilon} P[A\,m(Z, \nu^\epsilon)] \Bigm|_{\epsilon=0} &= P\Big[ A \, \partial_\nu m(Z) \times \frac{d}{d\epsilon} \nu(P^\epsilon) \Bigm|_{\epsilon=0}  \Big] = P [A \, \partial_\nu m(Z) \times \dot{\nu}(P;Q-P) ].
\end{align*}
It is then important to recall that $\dot{\nu}(P;Q-P)$ does not depend on $Z$, hence can be taken out of the expectation (in the conditional case, one will instead use the measurability property of conditional expectation). Equation \eqref{eq:diff-identity} thus becomes
\begin{align} \label{eq:diff-equiv}
	(Q-P)[A \, m(\nu)] + P[A \, \partial_\nu m] \, \dot{\nu}(P; Q-P) = 0.
\end{align}
Given that $P[A \, \partial_\nu m]$ is assumed to be invertible, we get
\begin{align*}
	\dot{\nu}(P; Q-P) = (Q-P) \big[ - \big(A \, P[\partial_\nu m]\big)^{-1} A \, m(Z, \nu) \big].
\end{align*}

Under the same invertible assumption, there exists a symmetric weighting matrix $\calW$ such that $A = P[\partial_\nu m]^\intercal \calW$. Since $m \in \mathcal{L}_2^0(P)$, the influence function is
\begin{align} \label{eq:IndirectIF}
	\dot{\nu}(Z; P) = - \big( P[\partial_\nu m]^\intercal \, \calW \, P[\partial_\nu m] \big)^{-1} P[\partial_\nu m]^\intercal \, \calW \, m \eqqcolon - P[\partial_\nu m]_{L}^{-1} \msone m(Z, \nu),
\end{align}
where $P[\partial_\nu m]_{L}^{-1}$ denotes the left inverse of $P[\partial_\nu m]$. Notably, the influence function is a linear transformation of the moment function, even though $m$ may be nonlinear in $\nu$. In the direct identification case $m(z,\nu) \!=\! h(z) - \nu$, this simplifies further: $m$ is linear in $\nu$ and $P[\partial_\nu m]_{L}^{-1} \!=\! -1$. This linear structure persists in the nonparametric setting (see Section~\ref{subsec:NonPar}).

A natural candidate for the optimal weighting matrix is the Moore--Penrose inverse $V_\nu^+$, where $V_\nu \coloneqq \langle m, m^\intercal \rangle$. The condition $(I - V_\nu V_\nu^+) P[\partial_\nu m] = 0$ is sufficient to ensure that this generalized inverse is indeed the optimal choice. When this condition holds, the efficient influence function writes as
\begin{align*}
	\dot{\nu}(Z;P)^\star = - \big( P[\partial_\nu m]^\intercal \, V_\nu^+ \, P[\partial_\nu m] \big)^{-1} P[\partial_\nu m]^\intercal \, V_\nu^+ \, m(Z,\nu).
\end{align*}


More generally, the moment function $m$ need not be classically differentiable with respect to $\nu$, in contrast to \cite{Hahn&Ridder:2013}, who impose standard smoothness conditions (see the statement preceding their Lemma~1). Instead, we employ distributional derivatives \citep{Jones:1982, Hormander:1983}, which extend differentiation to functions with discontinuities and justify the interchange of integration and differentiation.

\begin{example}[Non-differentiable moment] \label{exmp:non-smooth}
Consider the $q$-th quantile $\nu_q$ of $Z$, with $q \in (0,1)$. The moment function $m(z, \nu_q) \!=\! \mathds{1}_{\{z \leq \nu_q\}} - q$ is non-differentiable in the classical sense, as it corresponds to a shifted Heaviside function. Interpreting the derivative in the distributional sense, $\partial_{\nu_q} m$ corresponds to the Dirac delta $\delta(\nu_q - z)$, yielding $P[\partial_{\nu_q} m] \!=\! P[\delta(\nu_q - Z)] \!=\! \psk(\nu_q)$, where $\psk(\cdot)$ denotes the density of $Z$. Provided $\psk(\nu_q) \!>\! 0$, the influence function is a special case of \eqref{eq:IndirectIF}:
\begin{align*}
	\dot{\nu}_q(Z,P)  = - \Big( P \Big[ \frac{\partial m}{\partial \nu_q} \Big] \Big)^{-1}  m(Z, \nu_q) = - \frac{m(Z,\nu_q)}{\psk(\nu_q)} = \frac{q - \mathds{1}_{\{ Z \leq \nu_q \}}}{\psk(\nu_q)} .
\end{align*}
\end{example}

The quantile example illustrates a general principle: when the integrated map 
$\nu \mapsto P[g(Z,\nu)]$ is differentiable, the theory of distribution allows one to interchange the order of integration and differentiation (in the distributional sense) under much milder conditions than what the dominant convergence theorem requires. 
This mechanism extends beyond indicator functions to any moment condition where 
classical differentiability fails but the integrated map remains differentiable. The 
following lemma summarizes the above results into a more general chain rule.

\begin{lemma}[Chain Rule for Functional Derivatives: Unconditional Case] \label{lem:Key-Uncond}
Assume $\nu\!=\!\nu(P)$ is a regular parameter. Let $g(Z,\nu)$ be such that the integral map $\nu \mapsto P[g(Z,\nu)]$ is differentiable with respect to $\nu$. Then
\begin{align} \label{eq:chain-uncond}
	\frac{d}{d\epsilon} P\big[g(Z,\nu(P^\epsilon))\big] \Bigm|_{\epsilon=0} = \Big( \frac{\partial}{\partial \nu} P[g(Z,\nu)] \Big) \,\dot{\nu}(P; Q-P).
\end{align}
If we interpret $\partial_\nu g$ in the distributional sense (recall Example \ref{exmp:non-smooth}), the right-hand side can also be written as $P[\partial_\nu g(Z,\nu)]\,\dot{\nu}(P; Q-P)$.

Equivalently, let $\beta(P) \!\coloneqq\! P[g(Z, \nu(P))]$. Its directional and functional derivatives are given respectively by:
\begin{align} \label{eq:IF-chain-uncond}
\begin{split}
	\dot{\beta}(P; Q-P) &=  (Q-P)[g(Z,\nu)] + P\big[\partial_\nu g(Z,\nu)\big]\,\dot{\nu}(P; Q-P), \\
	\dot{\beta}(Z; P) &= \big( g(z, \nu) - P[g(Z, \nu)] \big) + P[\partial_\nu g(Z, \nu)] \, \dot{\nu}(Z; P).
\end{split}
\end{align}
\end{lemma}

To connect to the current indirect identification setting, we set 
$g(z,\nu) \!=\! A \, m(z, \nu)$. Since $\nu(P)$ satisfies the moment 
condition by definition, $\beta(P) \!=\! P[A \, m(Z, \nu(P))] \equiv 0$. 
Therefore, its directional derivative and influence function both vanish: 
$\dot{\beta}(P;Q-P)\!=\!\dot{\beta}(Z; P) \!\equiv\! 0$. Substituting into 
\eqref{eq:IF-chain-uncond} immediately recovers the linear system 
\eqref{eq:diff-equiv} used to solve for $\dot{\nu}(Z; P)$.

The introduction of $\beta(P)$ in Lemma \ref{lem:Key-Uncond}, however, serves a broader purpose beyond the current subsection: it mirrors the semiparametric plug-in estimation problem studied in Section \ref{sec:Semiparametric}. Imagine $\beta$ is the parameter of interest. Then we face a sequential estimation problem where one first estimates $\nu$ and then plugs it into $g$ to estimate $\beta$. 
In such a sequential problem, it is natural to see a composite functional of $P$: the distribution $P$ enters both directly, through the outer expectation, and indirectly, through the plug-in parameter $\nu(P)$. 

The structure of the influence function $\dot{\beta}(Z; P)$ in \eqref{eq:IF-chain-uncond} reflects this dual dependence. The first term, $g(Z, \nu) - P[g(Z, \nu)]$, would be the influence function of $\beta$ if $\nu$ were known (recall Section \ref{subsec:DirectInd}). The second term, $P[\partial_\nu g(Z, \nu)] \, \dot{\nu}(Z; P)$, captures the additional impact from estimating $\nu$. This structure underlies the analysis of sequential plug-in estimation, and Theorem \ref{thm:CondDirect} extends it to the semiparametric case.

Although Lemma \ref{lem:Key-Uncond} is stated for the unconditional case, it extends naturally to the conditional setting. The key requirement is whether the perturbed parameter is identified under the same distribution as the outer expectation. When this holds, the directional derivative can be factored out of the outer expectation by the measurability property of conditional expectation. We elaborate on this point in the subsequent sections and also discuss the important case where the two distributions differ.

\section{Nonparametric Influence Functions} \label{sec:Nonparametric}

\subsection{A simple setup} \label{subsec:NonPar}

We illustrate our method using the standard nonparametric regression model, where $Z \!=\! (X^\intercal, Y)^\intercal$ and the infinite-dimensional parameter is identified directly:
\begin{align*}
	Y = \gamma(X) + \epsilon,\quad \bbE[\epsilon|X] = P_{(Y|X)}[Y-\gamma(X)] = 0.
\end{align*}
A critical distinction must be drawn between $\gamma(X) \!=\! P_{(Y|X)}[Y]$, which depends on the random variable $X$, and its point evaluation $\gamma_x \!\coloneqq\! \gamma(x)$ at any given point $x$ in the support of $X$. 

Expanding the perturbed joint measure to first order in $\epsilon$ yields
\begin{align*}  
	dP^\epsilon &= dP_{(X)} dP_{(Y|X)} + \epsilon [dQ_{(X)} - dP_{(X)}] dP_{(Y|X)} + \epsilon \msone dQ_{(X)} [dQ_{(Y|X)} - dP_{(Y|X)}] \\
	&= dP_{(X)}^\epsilon dP_{(Y|X)} + \epsilon \msone dQ_{(X)} [dQ_{(Y|X)} - dP_{(Y|X)}].
\end{align*}
Algebraic rearrangement gives the corresponding perturbation of the conditional law:
\begin{align} \label{eq:CondPtbn}
	dP_{(Y|X)}^\epsilon = \frac{dP^\epsilon}{dP_{(X)}^{\epsilon}} = dP_{(Y|X)} + \epsilon (dQ_{(Y|X)} - dP_{(Y|X)}) \frac{dQ_{(X)}}{dP_{(X)}^\epsilon}.
\end{align}
Crucially, the identification $\gamma(X) \!=\! P_{(Y|X)}[Y]$ depends only on the conditional measure and is invariant to perturbations of the marginal $P_{(X)}$. This invariance permits a direct derivation of $\dot{\gamma}(P_{(Y|X)})$ (refer to the Proof of Theorem~\ref{thm:CondDirect} for a general treatment). Setting $Q_{(X)} \!=\! P_{(X)}$, which leads to $dQ_{(X)}/dP_{(X)}^\epsilon \!\equiv\! 1$, and extending Definition~\ref{def:derivative} to conditional measures yields
\begin{align} \label{eq:condIF}
\begin{gathered}
	\dot{\gamma}(P_{(Y|X)}; Q - P) = \dot{\gamma}(P_{(Y|X)}; Q_{(Y|X)} - P_{(Y|X)}) = ( Q_{(Y|X)} - P_{(Y|X)} )[Y] \\
\,\Longrightarrow\, \dot{\gamma}(Z, P_{(Y|X)}) = Y - P_{(Y|X)}[Y] = (Y- \gamma(X) ).
\end{gathered}
\end{align}
This mirrors the direct identification structure discussed in Section~\ref{subsec:DirectInd}. Under the standard assumption that $\Var(Y \!\mid\! X)$ is finite, the infinite-dimensional parameter $\gamma(\cdot)$ satisfies the regularity conditions outlined in Definition~\ref{def:regularity}.

\begin{example}[Propensity Score]
In treatment effect analysis, where $Z \!=\! (X,T,Y)$, the propensity score is directly identified as $\pi(X) \!\coloneqq\! P_{(T|X)}[T]$. Its influence function is simply $\dot{\pi}(Z, P_{(Y|X)}) = T - P_{(T|X)}[T] = T - \pi(X)$.
\end{example}

By contrast, the point evaluation $\gamma_x$ depends explicitly on the marginal density $\psk(x) \!=\! P[\delta_x(X)]$. Specifically, $\gamma_x$ is a nonlinear functional of both $P$ and the Dirac measure $\delta_x$:
\begin{align*}
	\gamma_x = \frac{P\big[P_{(Y|X)}[ \delta_x(X) Y] \big]}{P[\delta_x(X)]} = \frac{P[  \delta_x(X) Y]}{P[\delta_x(X)]} \eqcolon \gamma(P, \delta_x).
\end{align*}
Constructing an estimator for $\gamma_x$ requires approximating the singular functional $\delta_x$ with a sequence of regular functionals. Consider the Nadaraya–Watson (NW) kernel estimator and the linear sieve estimator with a growing basis $\boldsymbol{\phi}_J(X) = (\phi_1(X), \dots, \phi_J(X))^\intercal$:
\begin{align*}
	\hat{\gamma}^{\texttt{NW}}_x = \frac{\bbP_n \big[  k_b(x-X) Y \big]}{\bbP_n[ k_b(x-X)]} \,\, \text{ and } \,\, \hat{\gamma}^{\texttt{LSv}}_x = \boldsymbol{\phi}_J(x)^\intercal \big( \bbP_n[\boldsymbol{\phi}_J \boldsymbol{\phi}_J^\intercal ] \big)^{-1} \bbP_n[\boldsymbol{\phi}_J Y].
\end{align*}
The primary reason for employing these nonparametric methods is that their estimators admit closed-form expressions, which facilitates a transparent demonstration of our approach.

The approximation of $\delta_x$ induces asymptotic bias, whereas the substitution of $P$ with $\bbP_n$ governs the stochastic fluctuation. To formalize this decomposition, define a smoothed (or approximating) functional $\gamma^{\texttt{B}}_x(P)$ for each case, satisfying $\hat{\gamma}_x = \gamma_x^{\texttt{B}}(\bbP_n)$:
\begin{align*}
	\gamma^{\texttt{B}}_x(P) = \frac{P \big[  k_b(x-X) Y \big]}{P[ k_b(x-X)]} \,\, \text{ or } \,\, \gamma^{\texttt{B}}_x(P) = \boldsymbol{\phi}_J(x)^\intercal \big( P[\boldsymbol{\phi}_J \boldsymbol{\phi}_J^\intercal ] \big)^{-1} P[\boldsymbol{\phi}_J Y].
\end{align*}
This yields the standard bias–variance decomposition:
\begin{align} \label{eq:nonpar-decom}
	\hat{\gamma}_x - \gamma_x = \gamma_x^{\texttt{B}}(\bbP_n) - \gamma_x = \underbrace{\gamma_x^{\texttt{B}}(\bbP_n) - \gamma_x^{\texttt{B}}(P)}_{\text{influence}} + \underbrace{\gamma_x^{\texttt{B}}(P) - \gamma_x}_{\text{bias}}.
\end{align}
The bias component originates from the regularization (or approximation) of $\delta_x$, whereas the first term captures the sampling variability induced by the empirical measure. 

In the parametric setting, direct identification yields a functional that is linear in the underlying distribution (Section~\ref{subsec:DirectInd}). By contrast, the biased nonparametric functional $\gamma_x^{\texttt{B}}(P)$ is inherently nonlinear due to the normalization induced by kernel or sieve approximations. Nevertheless, the directional derivative $\dot{\gamma}_x^{\texttt{B}}(P; \bbP_n - P)$ continues to characterize the dominant stochastic term in the von Mises expansion of $\gamma_x^{\texttt{B}}(\bbP_n) - \gamma_x^{\texttt{B}}(P)$. Specifically, the derivative isolates the first-order empirical process fluctuation, while higher-order remainders are of smaller orders.

Applying a suitably modified version of Definition~\ref{def:derivative} to $\gamma_x^{\texttt{B}}$ (see below), the functional derivative $\dot{\gamma}_x^{\texttt{B}}(Z,P)$ serves as the nonparametric influence function. While its derivation requires careful handling of the approximation sequence, it yields transparent closed-form expressions that foreshadow the treatment of sequential plug-in semiparametric models. Below, we present two complementary approaches to unify the kernel and sieve specifications within our differentiation framework.

\subsection{Various calculation methods} \label{subsec:calculation}

In this subsection, we will discuss several ways of calculating the nonparametric influence function $\dot{\gamma}_x^{\texttt{B}}(Z,P)$, which are also helpful in the semiparametric case. 

\subsubsection{A nonlinear function perspective} 
The first approach is to view $\gamma^{\texttt{B}}_x$ as a special case of a generic parameter $\nu_x(P) \!=\! c^\intercal (P[g])^{-1} P[h]$:
\begin{align*}
\begin{dcases}
	\text{NW} & c=1, \, g(Z) = k_{b,x}(X), \, h(Z) = k_{b,x}(X) Y \\
	\text{LSv} & c = \boldsymbol{\phi}_J(x), \, g(Z) =\boldsymbol{\phi}_J(X) \boldsymbol{\phi}_J(X)^\intercal, \, h(Z) = \boldsymbol{\phi}_J(X) Y
\end{dcases}.
\end{align*}
Although the linear sieve case involves matrix-valued functions, the derivation of influence function remains rather intuitive. Applying the product rule to $M M^{-1} \!\equiv\! I$ (where $M$ is any invertible matrix-valued function) yields:
\begin{align*}
	dM \times M^{-1} + M \times dM^{-1} = 0 \;\; \Longrightarrow \;\; dM^{-1}= - M^{-1} (dM) M^{-1}.
\end{align*}
Based on this, we have
\begin{align*}
	\dot{\nu}_x(Z,P) &= c^\intercal (P[g])^{-1} (h-P[h])- c^\intercal (P[g])^{-1} (g-P[g]) (P[g])^{-1} P[h] \\
	&= c^\intercal (P[g])^{-1} \big( h - g \, (P[g])^{-1} P[h] \big).
\end{align*}
Plugging in the corresponding values for $c, g, h$, the nonparametric influence functions for NW and linear sieve estimators are
\begin{align} \label{eq:npIF}
	\dot{\gamma}^{\texttt{B}}_x(Z,P) = k_{b,x}^\dag(X,P) ( Y - \gamma^{\texttt{B}}_x(P) ) \;\; \text{ or } \;\; \dot{\gamma}^{\texttt{B}}_x(Z,P) = \boldsymbol{\phi}_{J,x}^\dag(X, P) (Y - \gamma^{\texttt{B}}_X(P) ),
\end{align}
where both $k_{b,x}^\dag(\cdot,P)$ and $\boldsymbol{\phi}_{J,x}^\dag(\cdot, P)$ are approximations of $\delta_x^\dag(\cdot,P) \!\coloneqq\! (P[\delta_x])^{-1} \delta_x(\cdot)$:
\begin{align} \label{eq:kphi}
	k_{b,x}^\dag(\cdot,P) \coloneqq (P[k_{b,x}])^{-1} k_{b,x}(\cdot) \,\,\text{and}\,\,
	\boldsymbol{\phi}_{J,x}^\dag(\cdot, P) \coloneqq \boldsymbol{\phi}_J(x)^\intercal (P[\boldsymbol{\phi}_J \boldsymbol{\phi}_J^\intercal])^{-1} \boldsymbol{\phi}_J(\cdot).
\end{align}

The structure of the influence function becomes clearer in this context, mirroring \eqref{eq:IndirectIF}, where $\dot{\nu}(Z,P)$ is linear in the moment function $m(Z, \nu)$. Here, the moment function is simply $m(Z, \gamma) \!=\! Y - \gamma(X)$, and $\dot{\gamma}^{\texttt{B}}_x$ is linear in $m(Z, \gamma^{\texttt{B}}_x)$. The ``loading'' term, such as $k_{b,x}^\dag$ or $\boldsymbol{\phi}_{J,x}^\dag$, reflects the estimation method and links the moment function to the influence function. Interestingly, the influence function links to $Y - \gamma^{\texttt{B}}_x(P)$ for the local nonparametric method and to $Y - \gamma^{\texttt{B}}_X(P)$ for the global method.

For fixed $b$ or $J$, the variance of $\dot{\nu}_x(P)$ is finite, but to ensure $\gamma^{\texttt{B}}_x - \gamma_x \!=\! o_p(1)$, we need $b \!\to\! 0$ or $J \!\to\! \infty$, leading to diverging variances in the loading terms. More specifically, let $t \!=\! b$ or $J$ be the tuning parameter and $f_{t,x}^\dag$ represents the nonparametric method. Denote by $r_t$ the divergence rate such that the limit of $P[f_{t,x}^\dag (f_{t,x}^\dag)^\intercal]/r_t^2$ is finite. Then the rate-adjust parameter $r_t^{-1} \gamma^{\texttt{B}}_x$ is regular. Thus, we can view $\gamma_x^{\texttt{B}}$ as a rate-adjusted regular parameter. Accordingly, its rate of convergence is $\sqrt{n} r_t^{-1}$, which is slower than the root-$n$ rate.

\subsubsection{A composite function perspective} 
Using the normalized weights defined in \eqref{eq:kphi}, we can express the estimators as composite functionals of $P$:
\begin{align*}
	\gamma_x = P[\delta_x^\dag(X,P) \msone  Y] \; \mapsto \; \hat{\gamma}^{\texttt{NW}}_x = \bbP_n[k_{b,x}^\dag(X,\bbP_n) \msone  Y] \text{ or }  \hat{\gamma}^{\texttt{LSv}}_x = \bbP_n[ \boldsymbol{\phi}_{J,x}^\dag(X,\bbP_n) \msone  Y].
\end{align*}

Accordingly, the smoothed functional $\gamma^{\texttt{B}}_x(P)$ is a special case of the composite functional:
\begin{align} \label{eq:nux}
	\nu_x(P) = P[ f_{t,x}^\dag(X, P) Y ], \text{ where, e.g., } f_{t,x}^\dag(X, P) = k_{b,x}^\dag(X, P) \text{ or } \boldsymbol{\phi}_{J,x}^\dag(X, P).
\end{align}
This representation is somewhat similar to the composite structure of the parameter $\beta(P) \!=\! P[g(Z, \nu(P))]$ in Lemma \ref{lem:Key-Uncond}, by viewing $f_{t,x}^\dag(X, P)$ as the counterpart of $\nu(P)$. However, the key distinction is that $f_{t,x}^\dag(X, P)$ depends on both $P$ and $X$, whereas $\nu(P)$ only depends on $P$.

Therefore, one can not directly apply Lemma \ref{lem:Key-Uncond} here but needs to go back to directional derivative. Specifically, we have
\begin{align*}
	\dot{\nu}_x(P; Q-P) &= (Q-P)[ f_{t,x}^\dag(X, P) Y ] + P[ \dot{f}_x^\dag(X,P; Q-P) Y ].
\end{align*}
The second term $P[ \dot{f}_x^\dag(X,P; Q-P) Y ]$ needs to be handled more carefully. Different from $\dot{\nu}(P;Q-P)$ in Lemma \ref{lem:Key-Uncond}, the term $\dot{f}_x^\dag(X,P; Q-P)$ depends on the random vector $X$, hence can not be simply taken out from the expectation sign as in the unconditional case.

Similar to the calculation in the previous subsection, we first obtain:
\begin{gather*}
 \dot{k}_{b,x}^\dag(X, P; Q-P) = - \frac{ k_{b,x}(X)}{P[k_{b,x}]^2} (Q-P)[ k_{b,x}(X)] = - k_{b,x}^\dag(X, P) \times (Q-P)[k_{b,x}^\dag(X, P)], \\
	\dot{\boldsymbol{\phi}}_{J,x}^\dag(X, P; Q-P) = - \boldsymbol{\phi}_J(x)^\intercal P[g]^{-1} \times (Q-P)[ \boldsymbol{\phi}_J(X) \boldsymbol{\phi}_J(X)^\intercal ] \times P[g]^{-1} \boldsymbol{\phi}_J(X).
\end{gather*}
In the NW specification, it is easy to see that ($\nu_x(P)=P[ k_{b,x}^\dag(X, P) Y ]$ is constant)
\begin{align*}
	& P[ \dot{k}_{b,x}^\dag(X,P_X; Q-P) Y] = -  P\{ k_{b,x}^\dag(X, P) Y \cdot (Q-P)[k_{b,x}^\dag(X, P)] \} \\
	=\,& - P[ k_{b,x}^\dag(X, P) Y ] \cdot (Q-P)[k_{b,x}^\dag(X, P)] = -  (Q-P)[k_{b,x}^\dag(X, P) \nu_x(P) ]. 
\end{align*}
For the linear sieve estimator, the coefficient $\boldsymbol{\phi}_J(x)^\intercal P[g]^{-1}$ is also nonrandom:
\begin{align*}
	& P[ \dot{\boldsymbol{\phi}}_{J,x}^\dag(X, P; Q-P) \msone Y] \\
	=\,& -\boldsymbol{\phi}_J(x)^\intercal P[g]^{-1}  (Q-P)[ \boldsymbol{\phi}_J(X) \boldsymbol{\phi}_J(X)^\intercal ]  P[g]^{-1} P[  \boldsymbol{\phi}_J(X) Y ] \\
	=\,& -(Q-P) \{ \boldsymbol{\phi}_J(x)^\intercal P[g]^{-1} \boldsymbol{\phi}_J(X) \times \boldsymbol{\phi}_J(X)^\intercal P[g]^{-1} P[  \boldsymbol{\phi}_J(X) Y ] \} \\
	=\,& -(Q-P) [ \boldsymbol{\phi}_{J,x}^\dag(X, P) \, \nu_X(P) ],
\end{align*}
where $\nu_X(P) \!=\! \boldsymbol{\phi}_J(X)^\intercal P[g]^{-1} P[  \boldsymbol{\phi}_J(X) Y ]$. Putting all together, one get the same results as in \eqref{eq:npIF}.

\subsection{A more general chain rule} 
Write $\gamma(P_{(Y|X)}) \!=\! P_{(Y|X)}[Y] \!=\! \gamma(X)$ to stress its dependence on $P_{(Y|X)}$. This yields a third representation of $\gamma_x(P)$ that mirrors the structure of semiparametric plug-in estimation in \eqref{eq:betaDirect}:
\begin{align*}
	\gamma_x^{\texttt{B}}(P) = P[ f_{t,x}^\dag(X, P) Y ] = P\big[ f_{t,x}^\dag(X, P) P_{(Y|X)}[Y] \big] = P[ f_{t,x}^\dag(X, P) \gamma(P_{(Y|X)}) ].
\end{align*}
If the weighting function $f_{t,x}^\dag(\cdot, P)$ were square-integrable rather than a regularized approximation to a Dirac delta function, this setup would correspond to a standard semiparametric model. Even in the current setting, we can multiply $f_{t,x}^\dag$ by $r_t^{-1}$ so that the adjusted weight becomes square-integrable even in the limit.

More generally, consider $r_t^{-1} \gamma_x^{\texttt{B}}(P)$ as a special case of the composite functional
\begin{align*}
	\beta_x(P) = P[g(x,Z,\eta(P),\gamma(P_{(Y|X)}))].
\end{align*}
The new parameter $\eta(P)$ captures the part of $r_t^{-1} f_{t,x}^\dag(X, P)$ that depends on $P$. For example, we have $\eta(P) \!=\! P[k_{b,x}]$ for the NW estimator and $\eta(P) \!=\! P[\boldsymbol{\phi}_J \boldsymbol{\phi}_J^\intercal]$ for the linear sieve estimator. The main generalization of this functional relative to $\beta(P) \!=\! P[g(Z, \nu(P))]$ in Lemma \ref{lem:Key-Uncond} is that it involves a statistical functional of the conditional distribution $P_{(Y|X)}$. 

The directional derivative writes as
\begin{align*}
	\dot{\beta}_x(P;Q-P) 
	=\,& (Q-P)\big[ g(x, Z, \eta(P), \gamma(P_{(Y|X)}) ) \big] + P[ \partial_\eta g ] \, \dot{\eta}(P; Q-P) \\
	& + \frac{d}{d\epsilon} P \big[ g(x,Z, \eta(P), \gamma(P_{(Y|X)}^\epsilon) ) \big] \Bigm|_{\epsilon=0}, 
\end{align*}
where the second term is a direct result of Lemma \ref{lem:Key-Uncond}. In the nonparametric regression context, the result is the same as $P[ \dot{f}_x^\dag(X,P; Q-P) Y ]$ calculated in the previous subsection. The main focus is the last term, for which we have the following lemma.

\begin{lemma}[Chain Rule for Functional Derivatives: Conditional Case] \label{lem:Key-Cond}
Let $Z \!=\! (X^\intercal,Y^\intercal)^\intercal$. Assume $\gamma(P_{(Y|X)})$ is a regular (infinite dimensional) parameter. Let $g(Z,\nu)$ be such that the conditional integral map $\nu \mapsto P_{(Y|X)}[g(x,Z,\eta,\nu)]$ is differentiable with respect to $\nu$. Then
\begin{align} \label{eq:chain-cond}
	\frac{d}{d\epsilon} P \big[ g(x, Z, \eta, \gamma(P_{(Y|X)}^\epsilon) ) \big] \Bigm|_{\epsilon=0} = (Q-P)\big[ P_{(Y|X)}[\partial_\nu g] \, \dot{\gamma}(Z, P_{(Y|X)}) \big],
\end{align}
where $\partial_\nu g$ is distributional derivative such that $P_{(Y|X)}[\partial_\nu g] = \partial_\nu P_{(Y|X)}[g]$. Accordingly,
\begin{align} \label{eq:IF-chain-cond}
\begin{split}
	\dot{\beta}_x(P;Q-P) =\,& (Q-P)[g] + (Q-P)\big[ P[ \partial_\eta g ] \, \dot{\eta}(Z, P) \big] \\
	& + (Q-P)\big[ P_{(Y|X)}[\partial_\nu g] \, \dot{\gamma}(Z, P_{(Y|X)}) \big], \\	
	\dot{\beta}_x(Z, P) =\,& g - P[g] + P[ \partial_\eta g ] \, \dot{\eta}(Z, P) + P_{(Y|X)}[\partial_\nu g] \, \dot{\gamma}(Z, P_{(Y|X)}).
\end{split}
\end{align}
\end{lemma}

The term $P_{(Y|X)}[\partial_\nu g]$ corresponds to $\alpha_0(X)$ in \cite{Newey:1994, Ichimura&Newey:2022, CNS:2024}. The proof is not long and reveals a key change-of-measure effect when handling perturbation of conditional distribution. We begin from the definition:
\begin{align*}
	& \frac{d}{d\epsilon} P \big[ g(x,Z,\eta, \gamma(P_{(Y|X)}^\epsilon) ) \big] \Bigm|_{\epsilon=0} \\
	=\,& \lim_{\epsilon\rightarrow0} \frac{1}{\epsilon} P[g(x,Z,\eta, \gamma(P_{(Y|X)}^\epsilon)) - g(x,Z,\eta, \gamma(P_{(Y|X)}) )] \\
	=\,& \lim_{\epsilon\rightarrow0} \frac{1}{\epsilon} P_{(X)} \big[ P_{(Y|X)}[g(x,Z,\eta, \gamma(P_{(Y|X)}^\epsilon)) - g(x,Z,\eta, \gamma(P_{(Y|X)}) )] \big] \\
	=\,& \lim_{\epsilon\rightarrow0} \Big( \frac{1}{\epsilon} P_{(X)}\big[ P_{(Y|X)}[\partial_\gamma g] \cdot (\gamma(P_{(Y|X)}^\epsilon) - \gamma(P_{(Y|X)})) \big] + o(\epsilon) \Big) \\
	=\,& \lim_{\epsilon\rightarrow0} \int P_{(Y|X=x)}[\partial_\gamma g] \, \frac{1}{\epsilon} (\gamma(P_{(Y|X=x)}^\epsilon) - \gamma(P_{(Y|X=x)})) dP_{(X)}(x) \\
	=\,&  \lim_{\epsilon\rightarrow0} \int P_{(Y|X=x)}[\partial_\gamma g] \Big( \int\int \dot{\gamma}(z, P_{(Y|X=x)}) d(Q_{(Y|X=x)}(y) - P_{(Y|X=x)}(y)) \\
	&\quad \times  \frac{dQ_{(X)}(x)}{dP_{(X)}^\epsilon(x)} \Big) dP_{(X)}(x) \;\; (\text{recall } \eqref{eq:CondPtbn}) \\ 
	=\,& \int P_{(Y|X=x)}[\partial_\gamma g] \Big( \int \dot{\gamma}(z, P_{(Y|X)}) d(Q_{(Y|X=x)}(y) - P_{(Y|X=x)}(y)) \Big) dQ_{(X)}(x) \\
	=\,& Q_{(X)} \big[ P_{(Y|X)}[\partial_\gamma g] \, (Q_{(Y|X)} - P_{(Y|X)})[\dot{\gamma}(Z, P_{(Y|X)})] \big].
\end{align*}
The limit
\begin{align*}
	\lim_{\epsilon\rightarrow0} \frac{dQ_{(X)}}{dP^\epsilon_{(X)}}  dP_{(X)} = d Q_{(X)}
\end{align*}
formalizes the change-of-measure effect: the outer integration with respect to $P_{(X)}$ is replaced by integration against $Q_{(X)}$.

The conditional zero-mean property $P_{(Y|X)}[\dot{\gamma}(P_{(Y|X)})] \!=\! 0$ then yields
\begin{align*}
	& Q_{(X)} \big[ P_{(Y|X)}[\partial_\nu g] \, (Q_{(Y|X)} - P_{(Y|X)})[\dot{\gamma}(Z, P_{(Y|X)})] \big] \\
	=\,& Q_{(X)} \big[ P_{(Y|X)}[\partial_\gamma g] \, Q_{(Y|X)} [\dot{\gamma}(Z, P_{(Y|X)})] \big] = Q_{(X)} \big\{ Q_{(Y|X)} \big[ P_{(Y|X)}[\partial_\gamma g] \, \dot{\gamma}(Z, P_{(Y|X)}) \big] \big\} \\
	=\,& Q\big[ P_{(Y|X)}[\partial_\gamma g] \, \dot{\gamma}(Z, P_{(Y|X)}) \big] = (Q-P)\big[ P_{(Y|X)}[\partial_\gamma g] \, \dot{\gamma}(Z, P_{(Y|X)}) \big].
\end{align*}
The results then readily follows.

In this particular proof, we use the distribution pair $P$ and $P_{(Y|X)}$. But a close examination shows that it can be easily extended to more general pairs such as $\Pon{j}$ and $\Pon{j'}$ with $j<j'$. In addition, the above lemma does not only apply to point evaluation of a function, such as $\gamma_x$, but also to the entire function $\gamma(\cdot)$. Below is an example that synthesizes the nonlinear and composite perspectives.

\begin{example}[Potential outcomes] \label{exmp:cond-outcome-1}
In the average treatment effect (ATE) context, the conditional potential outcomes are defined as
\begin{align*}
	\tau_1(X) \coloneqq P_{(Y(1) | X)}[Y(1)] \quad\text{and}\quad \tau_0(X) \coloneqq P_{(Y(0) | X)}[Y(0)].
\end{align*}

Under unconfoundedness, these reduce to observed conditional expectations:
\begin{align*}
	\tau_1(X) = P_{(Y|T=1, X)}[Y] = \frac{P_{(Y,T|X)}[TY]}{P_{(T|X)}[T]}, \quad \tau_0(X) = P_{(Y|T=0, X)}[Y] = \frac{P_{(Y,T|X)}[(1-T)Y]}{P_{(T|X)}[1-T]}.
\end{align*}
Both parameters can thus be treated as directly identified, but the fact that $P_{(T|X)}$ is part of $P_{(Y,T|X)}$ complicates the calculation. Define
\begin{align*}
	\gamma_Y=\Peg{Y|T,X}[Y],\,\, \gamma_{1T}  = \Peg{T|X}\big[ T \gamma_Y \big], \,\, \gamma_{0T}  = \Peg{T|X}\big[ (1-T) \gamma_Y \big].
\end{align*}
Hence, both $\gamma_{1T}$ and $\gamma_{0T}$ are composite functions of $P$. Besides, the ratio structure of $\tau_1 = \gamma_{1T} / \pi$ and $\tau_0 = \gamma_{0T} / (1-\pi)$ introduces further nonlinearity. 

Applying Lemma \ref{lem:Key-Cond} with $P$ and $P_{(Y|X)}$ replaced by $P_{(T|X)}$ (the same as $P_{(Y,T|X)}$) and $P_{(Y|T,X)}$ respectively, we obtain (inputs are omitted for simplicity)
\begin{gather*}
	\dot{\gamma}_{1T} = T \gamma_Y - P_{(T|X)}[T\gamma_Y] + P_{(Y|T,X)}[T] \dot{\gamma}_Y = T\gamma_Y - \gamma_{1T} + T (Y - \gamma_Y) = TY - \gamma_{1T}, \\
	\dot{\gamma}_{0T} = (1-T) \gamma_{Y} - P_{(T|X)}[(1-T)\gamma_Y] + P_{(Y|T,X)}[1-T] \dot{\gamma}_Y = (1-T) Y - \gamma_{0T}.
\end{gather*}
Together with the product rule, we get
\begin{align*}
	\dot{\tau}_1 &= - \frac{\gamma_{1T}}{\pi^2} \dot{\pi} + \frac{\dot{\gamma}_{1T} }{\pi} = - \frac{\gamma_{1T}}{\pi^2} (T-\pi) + \frac{TY-\gamma_{1T}}{\pi} \\
	&= \frac{1}{\pi^2} \big( - T \gamma_{1T} + \pi \gamma_{1T} + TY \pi - \pi \gamma_{1T} \big) = \frac{T}{\pi} (Y-\tau_1), \\
	\dot{\tau}_0 &= \frac{\gamma_{0T}}{(1-\pi)^2} \dot{\pi} + \frac{\dot{\gamma}_{0T} }{1-\pi} = \frac{1}{(1-\pi)^2} \big[ (1-T)(1-\pi) Y - (1-T) \gamma_{0T} \big] \\
	&= \frac{(1-T)}{1-\pi} (Y-\tau_0).
\end{align*}
In the next section, we will present a much easier way to calculate these two influence functions.


\end{example}

\subsection{More examples and discussions} 

\begin{example}[Functional coefficient] 
Consider the exogenous functional coefficient model with $Z^{(1)} \!=\! (W^\intercal,X^\intercal)^\intercal$ and $Z^{(2)} \!=\! Y$:
\begin{align*}
	Y = \beta(W)^\intercal X + \epsilon, \quad P_{(Y|W,X)}[Y - \beta(W)^\intercal X] = 0.
\end{align*}
The local linear estimator of $\nu_w \coloneqq (\beta(w)^\intercal, \vec(\partial_w \beta)^\intercal )^\intercal$ at a fixed $w$ is
\begin{gather*}
	\hat{\nu}_w = \nu_w(\bbP_n) = \bbP_n[f_w^\dag(W, X, \bbP_n) Y], \text{ where } f_w^\dag(W, X, P) = ( P[ \tilde{X}_w k_{b,w} \tilde{X}_w^\intercal ]  )^{-1} \tilde{X}_w k_{b,w}.
\end{gather*}
The term $\tilde{X}(w) \!=\! E_w \otimes X $ with $E_w \!=\! (1, (W - w)^\intercal)^\intercal$. Applying any of the calculation methods yields
\begin{align*}
	\dot{\nu}_w^{\texttt{B}}(P) = f_w^\dag(W, X, P) (Y - \beta(W)^\intercal X).
\end{align*}
Extending the basis $E_w$ yields the standard local polynomial estimator \citep{Fan&Gijbels:1996}.
\end{example}

\begin{example}[Endogeneity] 
Consider a nonparametric instrumental variable model with $Z^{(1)} \!=\! W$ and $Z^{(2)} \!=\! (X, Y)$:
\begin{align*}
	Y = \gamma(X) + \epsilon,\quad P_{(Y,X|W)}[Y-\gamma(X)] = P[\delta_w^\dag(W, P) (Y-\gamma(X))] = 0.
\end{align*}
This is a well-known ill-posed inverse problem (see, e.g., \cite{Newey&Powell:2003}, \cite{Hall&Horowitz:2005}, \cite{Che&Reiss:2011}, and \cite{DFFR:2011} among others). The sieve method can be viewed as a projection onto a subspace, which regularizes the problem by reducing its dimensionality. One can also add an explicit penalty term to further regularize the problem. In particular, a ridge-type penalty can yield a closed-form solution for the estimator $\hat{\gamma}_x \!=\! \gamma_x(\bbP_n) \!=\! \bbP_n[f_{t,x}^\dag(W, \bbP_n) Y]$ with
\begin{align*}
	f_{t,x}^\dag(W, P) &= \phi_J(x) (P[ \phi_J \psi_K^\intercal] (P[\psi_K \psi_K^\intercal])^{-1} P[\psi_K \phi_J^\intercal] + \lambda )^{-1} \\
	&\qquad \times P[\phi_J \psi_K^\intercal] (P[\psi_K \psi_K^\intercal])^{-1} \psi_K(W).
\end{align*}
Here, $\psi_K(W)$ denotes a growing sieve basis for the instruments. When $\lambda\!=\!0$, identification requires $K\!\geq\! J$. The biased/smoothed functional and its influence function are
\begin{align*}
	\gamma_x^{\texttt{B}}(P) = P[f_{t,x}^{\dag}(W, P) Y] \text{ and } \dot{\gamma}_x^{\texttt{B}}(Z,P) = f_{t,x}^{\dag}(W, P) (Y - \gamma^{\texttt{B}}(X)).
\end{align*}
Despite the added complexity from endogeneity, which is absorbed into the weighting function $f_{t,x}^\dag(W, P)$, the multiplicative structure of the influence function remains identical to the exogenous case. 
\end{example}

In multivariate settings, the curse of dimensionality primarily impacts the approximation of the singular functional $\delta_x^\dag$ by $f_{x}^\dag$. Structural restrictions (e.g., additivity or low-dimensional interaction kernels) are commonly imposed to mitigate this. While such constraints modify the explicit form of $f_{x}^\dag$, they preserve the multiplicative structure of the influence function in \eqref{eq:npIF}, which is dictated solely by the underlying identification condition in \eqref{eq:Momdelta}.

Alternative approximation schemes (e.g.~local polynomials, neural networks, or other machine learning estimators) generate distinct weighting functions $f_{x}^\dag$ that may depend on $P$ in a more complex manner. Even when closed-form expressions are unavailable, the influence function intuitively should retain the canonical form $\dot{\gamma}_x^{\texttt{B}} \!=\! f_{x}^\dag \msone \dot{\gamma}(P_{(Y,X|W)})$, where the general form of $\dot{\gamma}(P_{(Y,X|W)})$ can be found in \eqref{eq:dotgammamom} below. Consequently, deriving the influence function remains a systematic application of the chain rule and the above expectation-substitution rule. Quantifying the associated nonparametric bias, which requires precise control over the approximation error of $\delta_x^\dag$, involves technically demanding mathematical analysis and goes beyond the scope of the current paper.

\section{Semiparametric Multi-Step Plug-in Estimation} \label{sec:Semiparametric}


Lemma~\ref{lem:Key-Cond} provides a direct route to deriving influence functions in semiparametric settings. We first establish the result for directly identified parameters.

\begin{theorem} \label{thm:CondDirect} 
Assumption~\ref{asmp:strong} holds true.
Let $\beta$ be a regular finite-dimensional parameter identified via
\begin{align} \label{eq:betaDirect}
	\beta(P) = P\big[h_\beta\big(Z,\gamma_1(P), \gamma_2(\Pon{1}), \ldots, \gamma_l(\Pon{l-1}) \big)\big],
\end{align}
where $h_\beta$ is a known function and each $\gamma_j$ is an unknown functional of $Z^{(j-1:1)}$. For $j'<j$, the function $\gamma_{j}$ could be an input to $\gamma_{j'}$. The influence function of $\beta$ is (recall $\Pon{j-1} \!\coloneqq\! P_{(l:j|j-1:1)}$):
\begin{align} \label{eq:IF-cond}
	\dot{\beta}(Z, P) = h_\beta(Z) + \sum_{j=1}^{l} \Pon{j-1} [ \partial_{\gamma_j} h_\beta ] \times \dot{\gamma}_j(Z, \Pon{j-1}),
\end{align}
where $\partial_{\gamma_j} h_\beta$ is understood in the distributional sense. 


If each $\gamma_j$ is directly identified as $\gamma_j(\Pon{j-1}) \!=\! \Pon{j-1}[h_{\gamma,j}(Z^{(j:1)})]$ for a known function $h_{\gamma,j}$, then $\dot{\gamma}_j(Z, \Pon{j-1}) \!=\! h_{\gamma,j}(Z) - \Pon{j-1}[h_{\gamma,j}]$. 
If $\gamma_j$ is indirectly identified via a known moment function $m_{\gamma,j}$ satisfying $\Pon{j-1}[m_{\gamma,j}(Z, \gamma_j)] \!=\! 0$, then (following similar argument as in Section \ref{subsec:IndirectInd})
\begin{align} \label{eq:dotgammamom}
\begin{split}
	\dot{\gamma}_j &= - \big( \Pon{j-1}[\partial_{\gamma_j} m_{\gamma,j}]^\intercal \msone \calW_j \msone \Pon{j-1}[\partial_{\gamma_j} m_{\gamma,j}] \big)^{-1} \Pon{j-1}[\partial_{\gamma_j} m_{\gamma,j}]^\intercal \msone \calW_j \msone m_{\gamma,j} \\
	&\eqqcolon - \Pon{j-1}[\partial_{\gamma_j} m_{\gamma,j}]_{L}^{-1} \msone m_{\gamma,j},
\end{split}
\end{align}
where $\calW_j$ is a symmetric positive semi-definite matrix function of $Z^{(j-1:1)}$. We omit the inputs to $\dot{\gamma}_j$ to save space.
\end{theorem}

The influence functions in \eqref{eq:IF-chain-cond} and \eqref{eq:IF-cond} share a similar structure. The proof of \eqref{eq:IF-cond} hinges on the change-of-measure effect discussed in the previous section (refer to the appendix for more details). 

The derived expression depends exclusively on the influence functions $\dot{\gamma}_j$, which do not contain any information about the nonparametric method, compared to their regularized approximations $\dot{\gamma}_j^{\texttt{B}}$. Consequently, the influence function $\dot{\beta}$ is invariant to the specific nonparametric methods employed for the nuisance parameters. This invariance extends to the indirect identification case (Theorem~\ref{thm:CondMoment}) and aligns with the classic result \citep{Newey:1994, ACHL:2014}: provided nuisance estimators are consistent and converge at appropriate rates, the asymptotic variance of $\hat{\beta}$ remains unchanged regardless of the estimation method for $\gamma$. 

This property also clarifies the choice of our examples in the previous subsection. While complex nonparametric or machine learning estimators affect higher-order remainders and bias terms, they do not alter the first-order influence function. A detailed analysis of these higher-order properties lies beyond the scope of this paper.

\begin{example}[Potential outcomes continued]
Recall Example \ref{exmp:cond-outcome-1} on potential outcomes. An easier way to calculate their influence functions is to use \eqref{eq:dotgammamom}. First note that the potential outcomes can also be identified respectively by $P_{(Y,T|X)}[m_{\tau_1}] \!=\! 0$ and $P_{(Y,T|X)}[m_{\tau_0}] \!=\! 0$, where
\begin{align} \label{eq:pitaumoments}
	m_{\tau_1} = TY - T \tau_1(X) \quad m_{\tau_0} = (1-T)(Y - \tau_0(X)).
\end{align}
Under standard overlap conditions, the following terms are almost surely invertible:
\begin{align*}
	P_{(Y,T|X)}[\partial_{\tau_1} m_{\tau_1}] = \pi(X), \quad P_{(Y,T|X)}[\partial_{\tau_0} m_{\tau_0}] = 1- \pi(X).
\end{align*}
Applying \eqref{eq:dotgammamom} yields
\begin{align} \label{eq:psoutcome}
	\dot{\tau}_1 = \frac{T}{\pi(X)} (Y - \tau_1(X)), \quad \dot{\tau}_0 = \frac{1-T}{1- \pi(X)} (Y - \tau_0(X)).
\end{align}
In this alternative approach, we avoid the explicit ratio structure $\tau_1 \!=\! \gamma_{1T}/\pi$ by turning it into a moment structure, thereby simplifying the calculation. 

\end{example}

\begin{example}[Average Treatment Effect (ATE)]
Under unconfoundedness, the identifying functionals for the inverse probability weighting (IPW) and regression-based estimators are 
\begin{align*}
	h_{\ATE}^{\IPW} = \frac{TY}{\pi(X)} - \frac{(1-T)Y}{1-\pi(X)}   \,\,\,\text{and}\,\,\,
	h_{\ATE}^{\Reg} = \tau_1(X) - \tau_0(X).
\end{align*}
Direct calculation shows that
\begin{gather*}
	P_{(Y,T|X)}[\partial_{\tau_1} h_{\ATE}^{\Reg}] = 1, \quad P_{(Y,T|X)}[\partial_{\tau_0} h_{\ATE}^{\Reg}] = -1, \\
	P_{(Y,T|X)} [\partial_\pi  h_{\ATE}^{\IPW}] = - \frac{\tau_1(X)}{\pi(X)} - \frac{\tau_0(X)}{1-\pi(X)}.
\end{gather*}
Substituting these into \eqref{eq:IF-cond} and using \eqref{eq:psoutcome} yields
\begin{align*}
	& \dot{\tau}_{\text{IPW}} = \dot{\tau}_{\text{Reg}} = \dot{\tau}_1 - \dot{\tau}_0 + \tau_1(X) - \tau_0(X) - \tau \\
	=\,& \frac{T}{\pi(X)} (Y - \tau_1(X)) - \frac{1-T}{1- \pi(X)} (Y - \tau_0(X)) + \tau_1(X) - \tau_0(X) - \tau.
\end{align*}
This expression coincides with the influence function of the doubly robust (AIPW) estimator.
\end{example}

Existing approaches for semiparametric moment models \citep{Ai&Chen:2003, Ichimura&Newey:2022} typically derive influence functions by solving abstract projection or minimum-distance/least-squares problems. Our differentiation-based framework bypasses these steps, yielding closed-form expressions without explicit tangent-space characterizations. The following theorem extends the result to indirect identification.

\begin{theorem} \label{thm:CondMoment}
Suppose the conditions of Theorem~\ref{thm:CondDirect} hold, except that $\beta$ satisfies the unconditional moment condition
\begin{align} \label{eq:Momentbeta}
	P\big[m_\beta\big(\beta, \gamma_1(P), \gamma_2(\Pon{1}), \ldots, \gamma_l(\Pon{l-1}) \big)\big] = 0, 
\end{align}
and let $\calW_{\beta\beta}$ be a conformable symmetric positive semi-definite matrix. The influence function of $\beta$ is (inputs are omitted to save space)
\begin{align} \label{eq:IF-ConM}
	\dot{\beta} = \big\{ P[\partial_\beta m_\beta]^\intercal \calW_{\beta\beta} \, P[\partial_\beta m_\beta] \big\}^{-1} P[\partial_\beta m_\beta]^\intercal \calW_{\beta\beta} \Big( m_\beta + \sum_{j=1}^l  \Pon{j-1} [\partial_{\gamma_j} m_\beta ] \dot{\gamma}_j \Big).
\end{align}
\end{theorem}

The expression enclosed in parentheses in \eqref{eq:IF-ConM} is precisely the locally robust moment function \citep{LocalRobust:2022}:
\begin{align*} 
	m^{\LR}_\beta = m_\beta + \sum_{j=1}^{l} \Pon{j-1} [\partial_{\gamma_j} m_\beta ] \times \dot{\gamma}_j.
\end{align*}
The asymptotic variance of $\dot{\beta}$ depends jointly on the nuisance weighting functions $\calW_j$ in \eqref{eq:dotgammamom} and the target weighting matrix $\calW_{\beta\beta}$. This raises a central efficiency question: when does the choice of $\calW_j$ that optimally estimates the nuisance parameters $\gamma$ also minimize the asymptotic variance of $\dot{\beta}$? We will leave this for future exploration.

The term $v_\rho$ in \cite{Ichimura&Newey:2022} corresponds to $\Pon{j-1}[\partial_{\gamma_j} m_{\gamma,j}]$ in \eqref{eq:dotgammamom}. However, their least-squares formulation relies on the ratio $-v_{m}(X)/v_\rho(X)$, which suggests that $v_\rho$ is scalar-valued. 

Our method yields the explicit results directly, without having to solve any other problem. Besides, it applies to vector- or even matrix-valued parameters. In the linear sieve example, the nuisance parameter $\eta(P) \!=\! P[\boldsymbol{\phi}_J \boldsymbol{\phi}_J^\intercal]$ is matrix-valued. In such cases, one just need to pick the representation of matrix calculus that preserves the chain rule \citep{Kollo&vonRosen:2006}.

\section{Conclusion} \label{sec:conclusion}

In this paper, we introduce a direct differentiation-based framework for deriving influence functions across parametric, nonparametric, and semiparametric models. By grounding the derivation in the functional derivative and its integral representation, we demonstrate that the influence function is obtained by centering the identification function, that is, subtracting its conditional or unconditional expectation. 

This algebraic characterization replaces the traditional reliance on tangent-space projections, score-based Riesz representation arguments, and ad-hoc conjecture-and-verify constructions. We formalize this insight through a unified framework that seamlessly handles both unconditional and conditional perturbations, yielding closed-form influence functions across parametric, nonparametric, and semiparametric cases.

\bigskip
\bigskip
\bigskip

\bibliographystyle{ecta}
\bibliography{YangBib}

\end{document}